\documentclass{emulateapj}

\shorttitle{Line broadening in coronal dimmings}
\shortauthors{L. Dolla \& A. Zhukov}

\usepackage{natbib}
\usepackage{color}
\usepackage{url}

\begin{document}

\title{On the nature of spectral line broadening in solar coronal dimmings}

\author{L. R. Dolla\altaffilmark{1} and A. N. Zhukov\altaffilmark{1,2}}
\email{dolla@sidc.be}

\altaffiltext{1}{Solar-Terrestrial Center of Excellence-SIDC, Royal Observatory of Belgium,
Avenue Circulaire 3, B-1180 Brussels, Belgium }
\altaffiltext{2}{Skobeltsyn Institute of Nuclear Physics, Moscow State University, 119992 Moscow, Russia }

\begin{abstract}
We analyze the profiles of iron emission lines observed in solar coronal dimmings associated with coronal mass ejections,  
using the EUV Imaging Spectrometer on board \textit{Hinode}. We quantify line profile distortions with empirical coefficients (asymmetry and peakedness) that compare the fitted Gaussian to the data. We find that the apparent line broadenings reported in previous studies are likely to be caused by inhomogeneities of flow velocities along the line of sight, or at scales smaller than the resolution scale, or by velocity fluctuations during the exposure time. The increase in the amplitude of Alfv\'en waves cannot, alone, explain the observed features.  
A double-Gaussian fit of the line profiles shows that, both for dimmings and active region loops, one component is nearly at rest while the second component presents a larger Doppler shift than that derived from a single-Gaussian fit. 
\end{abstract}


\maketitle

%
\section{Introduction}   
Coronal mass ejections (CMEs) are one of the most powerful energy release phenomena in the solar atmosphere. After such events, coronal dimmings usually appear as regions of temporary lowered intensity in coronal lines, near the erupting active region. 
The fainter emission is interpreted in terms of density decrease, as it is usually observed in several temperature regimes, and is usually linked to the coronal mass removal that follows the eruption \citep{Sterling97, Harrison03, Zhukov04}.  
Spectroscopic studies showed that coronal dimmings are associated with large outflows with speeds of tens of kilometers per second lasting several hours \citep{Harra01, Harra07}.
Using the EUV Imaging Spectrometer \citep[EIS, ][]{Culhane07} on board the \textit{Hinode} spacecraft, many observations revealed that strong outflows occurring in low-intensity regions were also associated with larger line widths than in the surrounding area \citep[e.g.][]{Delzanna08b}. 
\citet{Doschek08} suggest that these wide profiles may be due to turbulence or to the addition of several components having different intrinsic velocities. 
They even note that some profiles show secondary components that make the profile depart from a pure Gaussian profile. 
According to \citet{McIntosh09} and \citet{McIntosh09b}, the broadening is due to the growth of Alfv\'en wave amplitude in the magnetically open and rarefied region of a dimming. 
All these studies stress the possible connection between the solar wind and the observed outflows.  

From a theoretical point of view, both interpretations are possible. 
The corona is optically thin for most coronal emission lines. The observed lines are then the result of integration along the line of sight (LOS) and on scales smaller than the spatial resolution scales. This includes different layers having temperatures close to that of maximum formation of the line and also different magnetic flux tubes having differing plasma parameters. But integration is also done over the exposure time. 
Therefore, Alfv\'en waves and inhomogeneities of flows both produce spatial and temporal variations of Doppler velocities, which results in an apparent broadening of the integrated profile. 
The purpose of this paper is to investigate the distortion of the line profiles to distinguish between both interpretations. 
%
\section{Estimating the distortion of line profiles}   \label{sec estimating line profiles}
%
Different methods can be used to quantify the presence of additional components in line profiles. For example, \citet{Imada08} use a coefficient they call ``additional component contribution to the line broadening'', that makes use of the coefficients of a two-Gaussian fit that they apply in cases where the reduced chi-square ($\chi^2$) of the single-component fit is larger than $4.5$. 
\citet{McIntosh09c} evaluate the asymmetry by first finding the center of the line with a Gaussian fit, then calculating the difference of counts in symmetric intervals in the red and blue wings, for a given offset velocity interval. 

The classical skewness and kurtosis of a statistical distribution are influenced too much by discretization effects in the line profiles we observe, while we prefer not to interpolate into the profiles to avoid adding information that would not be present before. 
They nevertheless inspired us to propose two empirical coefficients that we call hereafter ``asymmetry'' and ``peakedness'' to avoid confusion. 
They enable us to analyze the distortion of line profiles when the separation between the components is hardly noticeable, for cases where $\chi^2 \approx 1$ with a single-Gaussian fit. 
As for a $\chi^2$ computation, we use the squared differences between the real data and a fitted Gaussian.  
But contributions are positive or negative according to the sign of the difference and the data point position on predefined intervals of the line profile. 

Each coefficient $C$ is defined as follows:
\begin{eqnarray} \label{eq coeff}
C & = & \frac{1}{N} \sum_{k} \epsilon(\lambda) \, \textrm{sgn}\left(s_k(\lambda)-f_k(\lambda) \right) \nonumber\\
 & & {}\times \left(\frac{s_k(\lambda)-f_k(\lambda)}{\varsigma_k(\lambda)}\right)^2, 
\end{eqnarray}
where $N$ is the total number of points where the contribution factor $\epsilon(\lambda)$, defined below, is non-zero.  The spectrum $s_k(\lambda)$ at wavelength $\lambda$ is discretized on spectral pixel $k$, $f_k(\lambda)$ is the fit to $s_k(\lambda)$, and $\varsigma_k(\lambda)$ is the error on $s_k(\lambda)$. For the asymmetry, $\epsilon(\lambda)$ is defined as follows, with $\lambda_0$ being the center and $\sigma$ being the half-width at $1/ \sqrt{e}$ of the fitted Gaussian:
\begin{eqnarray} \label{Eq intervals asym}
\epsilon(\lambda) = \left\{ \begin{array}{rl}
-1 & \textrm{ if } \lambda \in [\lambda_0 - 2 \, \sigma; \lambda_0 - \sigma) \\
1 & \textrm{ if } \lambda \in [\lambda_0 - \sigma; \lambda_0) \\
-1 & \textrm{ if } \lambda \in{} (\lambda_0; \lambda_0 + \sigma]\\
1 & \textrm{ if } \lambda \in{} (\lambda_0 + \sigma; \lambda_0 + 2 \, \sigma]\\
0 & \textrm{ otherwise. }
\end{array} \right.
\end{eqnarray}
For the peakedness:
\begin{eqnarray} \label{Eq intervals peak}
\epsilon(\lambda) = \left\{ \begin{array}{rl}
-1 & \textrm{ if } \lambda \in [\lambda_0 - 1.5 \, \sigma; \lambda_0 - 0.5 \, \sigma]\\
1 & \textrm{ if } \lambda \in{} (\lambda_0 - 0.5 \, \sigma; \lambda_0 + 0.5 \, \sigma) \\
-1 & \textrm{ if } \lambda \in [\lambda_0 + 0.5 \, \sigma; \lambda_0 + 1.5 \, \sigma]\\
0 & \textrm{ otherwise. }
\end{array} \right.
\end{eqnarray}
 
Tails beyond $2$ and $1.5 \sigma$, respectively, are not taken into account because they are more sensitive to noise. In practice, a coefficient equal to $1$ means that over the main part of the profile the deviation was on average equal to one error bar, but systematically above or below the fitted Gaussian according to the above defined intervals. This can then hardly be a coincidence and appears as a significant distortion. 

The basic idea is that any departure from a Gaussian profile may be due to the presence of additional components having different centers, widths, or amplitudes at the line center. 
For simplicity, we further show synthetic profiles compound of only two components and fitted them with single Gaussians. But the results can be extrapolated to the addition of several components, with the limitation that any distortion can be smoothed out when their number increases. As a simple example, we simulated spectra with a dominant component having an amplitude at the line center of 100 in photon units, varying the relative amplitude of the other component. Artificial error bars are used for the consistency of the definition of the coefficients, using Poisson statistic (note that the coefficients become less sensitive as the statistics decrease, because of the increased error bars). But for the basic behaviors analyzed here, no real noise is added: it can create an artificial distortion, or on the contrary may blur the real one, and requires statistical analysis beyond the scope of this paper. 

First, we study the case of two components having identical widths $\sigma$ (taken equal to 1.4 pixel, the average one measured in our data set, see below), with separations of the centers varying from 0 to $2 \, \sigma$; above $2 \, \sigma$, the double component becomes clearly visible. The coefficient of asymmetry increases (in an absolute value) when the separation between both lines increases, but reaches its maximum when the ratio between their amplitudes is around 50\%. This is easily understandable: if it is too weak, the additional component cannot significantly distort the profile, and if both amplitudes are too similar, the profile is only broadened without showing any asymmetry. 

\begin{figure*}
\begin{center}
\includegraphics[width=0.95\linewidth]{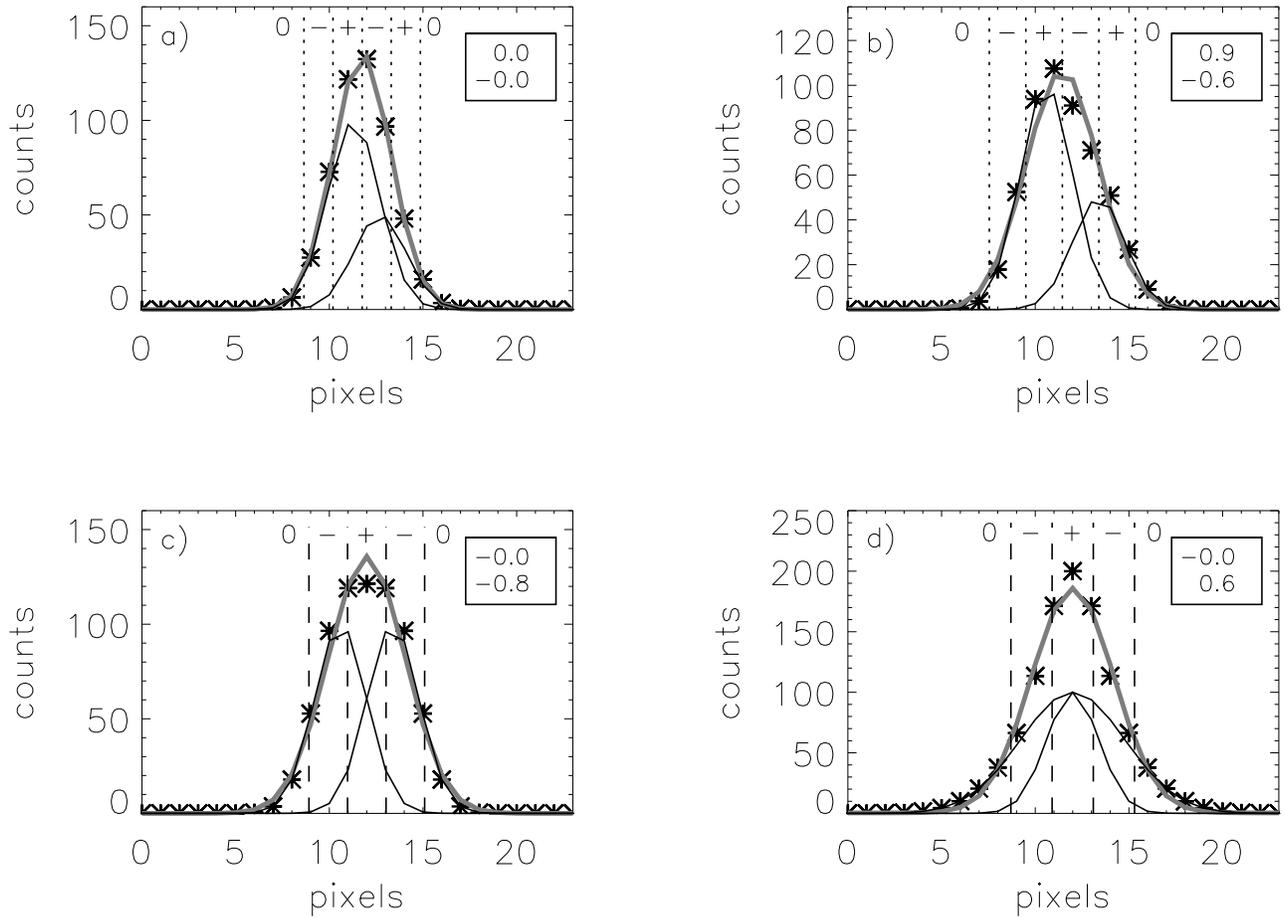}
\end{center}
\caption{Simulated profiles (asterisks) compound of two Gaussians (black solid lines) and fitted with a single Gaussian (gray solid line). \textit{(a)}: Sum of two Gaussians with the same width $\sigma = 1.4~\textrm{pixels}$, and centers separated by $1 \sigma$. The first one has an amplitude of 100 counts, and the second of 50 counts. \textit{(b)}: same as \textit{a)}, but with a separation of $2 \sigma$. \textit{(c)}: same as panel \textit{b)}, but the two Gaussians now have the same amplitude of 100 counts. \textit{(d)}: sum of two Gaussians of the same amplitude and same center, but widths of 1.4 and 2.8~pixels, respectively. 
The values in the insets are the coefficients of asymmetry and peakedness,  respectively. The ``$0$'', ``$+$'' and ``$-$'' on top of the curves indicate the sign of the contribution $\epsilon(\lambda)$ in the intervals delimited by the vertical lines; dotted lines for the intervals associated with the coefficient of asymmetry (cf Eq.~\ref{Eq intervals asym}) and dashed lines for that of the peakedness (cf Eq.~\ref{Eq intervals peak}; note that we separately represent the intervals for the asymmetry and the peakedness in the first and the second row, respectively, for better visibility). Points that are more than $2 \sigma$ away from the line center are then not taken into account. 
\label{fig simulated profiles}}
\end{figure*}

The upper panels of Fig.~\ref{fig simulated profiles} show two examples for a second component of 50\% amplitude, with a separation equal to $\sigma$ in panel \emph{(a)} and $2 \, \sigma$ in panel \emph{(b)}. In the case of panel  \emph{(a)}, it is impossible to say by eye that there are two components, and the coefficients of asymmetry and peakedness (shown in the inset of each panel) are roughly equal to 0 that is the value for a pure Gaussian (the first significant digit, not shown, is the third one after the decimal point). 
Nevertheless, the width of the resulting profile is already increased by 10\% as compared to that of the components. In the case of panel \emph{(b)}, the asymmetry is difficult to notice without the visual comparison with the fit. 
These two examples show that the fit tends to balance between the major component and the small tail by shifting the center of the fitted Gaussian from the major component toward the smaller one. The coefficients in Eqs.~(\ref{eq coeff}) to (\ref{Eq intervals peak}) were exactly designed to emphasize this behavior.  Note that while $\chi^2$ only estimates the deviation from the fitted Gaussian, the coefficient of asymmetry provides more refined information on the line distortion. Indeed, its sign gives the side on which the asymmetry is present: if positive, the profile has a major component on the shorter wavelengths (hence a tail at larger wavelengths), and vice versa. It can also detect asymmetries for the cases of $\chi^2$ values that would not be regarded as abnormal, i.e., close to 1. 

The coefficient of peakedness is less useful, because it is also influenced by the asymmetry of the profile. Nevertheless, it is noticeably negative (i.e., the profile is flatter than a Gaussian) when the components are similar in intensity and separated by more than $1 \sigma$ (flat-top profile, Fig.~\ref{fig simulated profiles} \textit{(c)}). Therefore, it is sometimes possible to detect the presence of two (or more) components when the asymmetry fails to do so. The peakedness is positive (i.e. the profile is more peaked) when the small component is less than 40\% in amplitude than the major one, still for large separations. It is also positive when both components have different widths (Fig.~\ref{fig simulated profiles} \textit{(d)}). In this last case, as long as the components have the same center, the asymmetry stays around 0. 
%
\section{Observations and data processing}   \label{sec obs and data processing}
%
\begin{deluxetable*}{crcccl}
\tablecaption{EIS data sets used in this study.  \label{tab EIS data sets}}
\tablewidth{0pt}
\tablehead{
\colhead{No.} & \colhead{Date} & \colhead{Start time\tablenotemark{a}} & \colhead{End time} & \colhead{$\tau$\tablenotemark{b}} & \colhead{Comment}
}
\startdata
1a & 2006 Dec 12 & 01:07:20 & 05:40:31 & 30 & \\
\phn b &          12 & 19:07:20 & 23:45:45 & 30 & After a B7.7 flare (16:45\tablenotemark{c})\\
\phn c &          13 & 01:12:12 & 05:41:09 & 30 & X3.4 flare at 02:14\\[1ex]
2a & 2006 Dec 14 & 19:20:12 & 21:34:24 & 30 & \\
\phn b &          15 & 01:15:19 & 03:29:31 & 30 & After an X1.5 flare (21:07 on December 14)\\
\phn c &          15 & 04:10:12 & 06:24:24 & 30 & Dimming partial recovery \\[1ex]
3a & 2007 Aug 22 & 13:33:46 & 17:55:56 & 60 & After a B1.2 flare (11:29)\\ 
\phn b &          23 & 01:55:43 & 06:17:53 & 60 & Overlap with an eruption\\
\enddata
\tablenotetext{a}{All times are UT. }
\tablenotetext{b}{Exposure time is in seconds. }
\tablenotetext{c}{All the flare times are flare start times as measured by \emph{GOES}. }
\end{deluxetable*}
We selected three \textit{Hinode}/EIS data sets that show coronal dimmings associated with solar eruptive events (Table~\ref{tab EIS data sets}). We first concentrate on the \ion{Fe}{12} 195.12~\AA{} line because its high signal is suitable for the study of line profiles. 
%
\subsection{Data set 1 : 2006 December 12-13}
%
In data set~1, NOAA AR 10930 was observed on 2006 December 12-13 with exposure times of 30~s for every Y-position. 
This region presents bright structures in the first EIS raster (set~1a, see the first column in Fig.~\ref{fig 13 dec}) diverging roughly from $\textrm{X}=-120\arcsec{}, \textrm{Y}=-120\arcsec{}$. Some field lines are clearly connected to the core of the active region. 
There are no data from the Extreme ultraviolet Imaging Telescope \citep[EIT,][]{Delaboudiniere95} on board the Solar and Heliospheric Observatory \citep[SOHO,][]{Domingo95} before 23:49:26 on December 12, but our next EIS raster (1b, second column in Fig.~\ref{fig 13 dec}) shows that this area presents a dimmed region a couple of hours after the \emph{GOES} B7.7 flare has started at 16:45 UT. An X3.4 flare then occurred at 02:14 UT and was also associated to the dimming in the same area. This event was observed by EIT. 
The EIS slit started to raster this dimming area one hour after the flare began \citep[set~1c, see Fig.~\ref{fig 13 dec main}, also analyzed by][]{Asai08}. All rasters are scanned from west to east. 
For the Doppler reference wavelength in raster~1c, we only used the southern ``quiet'' part (pixel rows 15 to 40) to reduce biases due to the large Doppler shifts present on most of the map.  
%
%
\subsection{Data set 2 : 2006 December 14-15}
%
Data set~2 corresponds to the same NOAA AR 10930 observed several hours after data set~1, again with 30~s exposure times. Set~2a (first column in Fig.~\ref{fig 14 15 dec}) shows that the dimmings have recovered since the X3.4 erupting flare. An X1.5 flare started at 21:07 UT on December 14. 
As noted by  \citet{Harra07} and \citet{McIntosh09}, a new dimming appears at the same location as during the December 13 event (second column, set~2b). Such a kind of repetition was described by \citet{Chertok04}. 
After a few hours, the dimming started to recover (set~2c). 
%
\subsection{Data set 3 : 2007 August 22-23}
%
Data set~3 (Fig.~\ref{fig 22 23 aug}) corresponds to an active region observed on 2007 August 22 and 23, also analyzed by \citet{Doschek08}. For these two rasters, the exposure times were 60~s. There is a B1.2 flare 2~hr  before the raster~3a, associated with dimmings that are best seen on images taken by the Extreme Ultraviolet Imager (EUVI), which is a part of the Sun Earth Connection Coronal and Heliospheric Investigation (SECCHI) suite
\citep{Howard08} on board the \textit{Solar TErrestrial RElations Observatory} (STEREO) mission \citep{Kaiser08}. 
The difference images in Fig.~\ref{fig diff images 22 aug} show a dimming in the southeast part of the active region, and a bright fanning-out structure that is preserved all along the observations. 

A second event occurs just before the beginning of the second raster (data set~3b). A faint dimming appears and the loop system clearly produces an eruption from 01:46 UT to 02:46 UT (Fig.~\ref{fig diff images 23 aug}), spanning the fanning-out structure. Around 03:56 UT (last panel), the scan is in the middle of the raster and the dimming area does not increase any more (with even some faint recovery), while bright post-eruptive loops have started to be seen in the active region core. There is no recorded flare in the \emph{GOES} catalog corresponding to this event, although one can observe a brightening in the \emph{GOES} light curve around 3:00, that corresponds to the appearance of the bright loops in the EUVI images.  

All three data sets exhibit the same properties that will be discussed in Sect.~\ref{sec results}. 
\begin{figure*}
\begin{center}
\includegraphics[height=0.95\textheight]{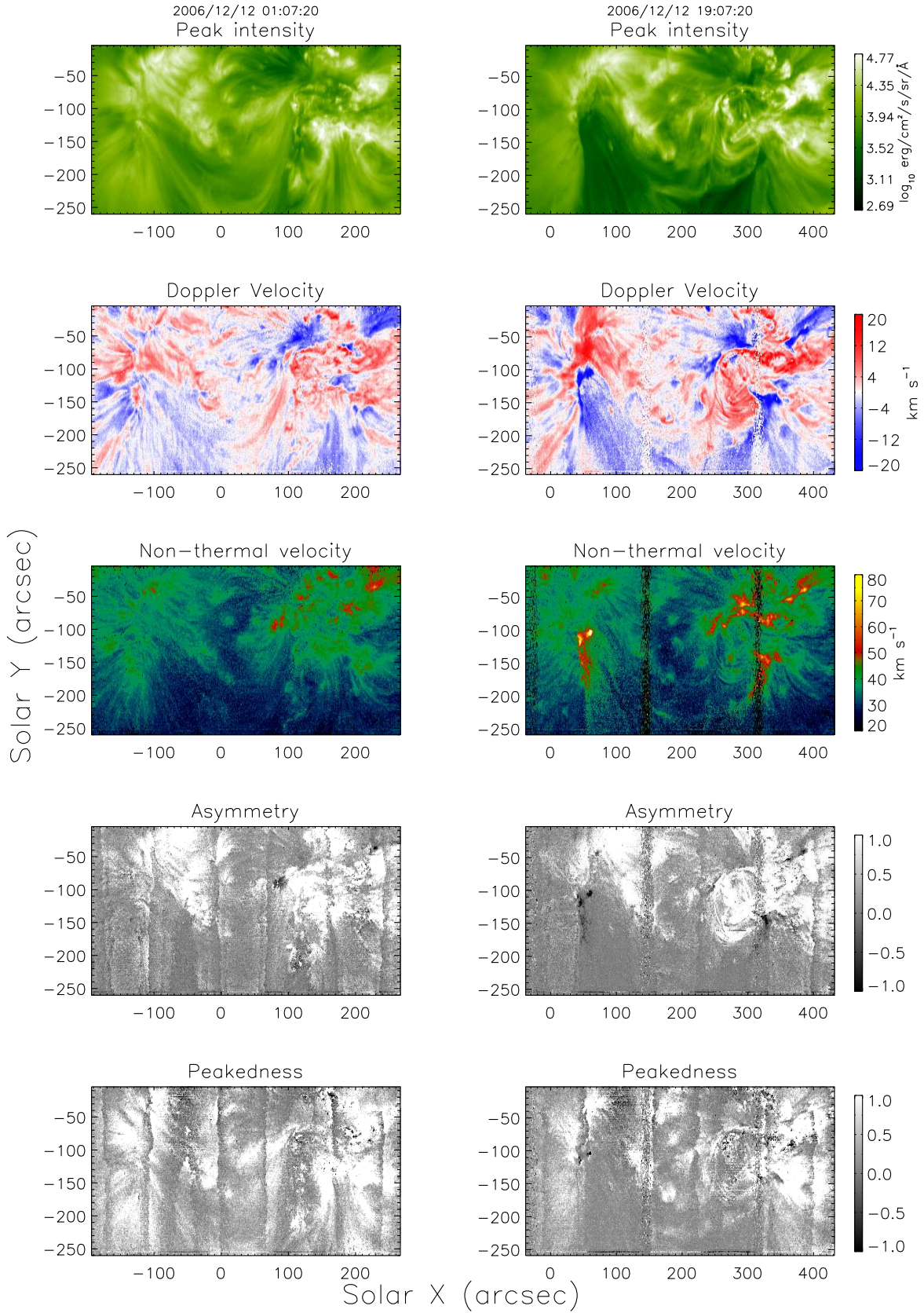}
\end{center}
\caption{From top to bottom: \ion{Fe}{12} 195.12~\AA{} line peak intensity, Doppler velocity, and non-thermal velocity observed by \textit{Hinode}/EIS on 2006 December 12 and 13 (data sets~1a and 1b). Also shown are the derived coefficients of asymmetry and peakedness. 
\label{fig 13 dec}}
\end{figure*}
\begin{figure}
\begin{center}
\includegraphics[height=0.94\textheight]{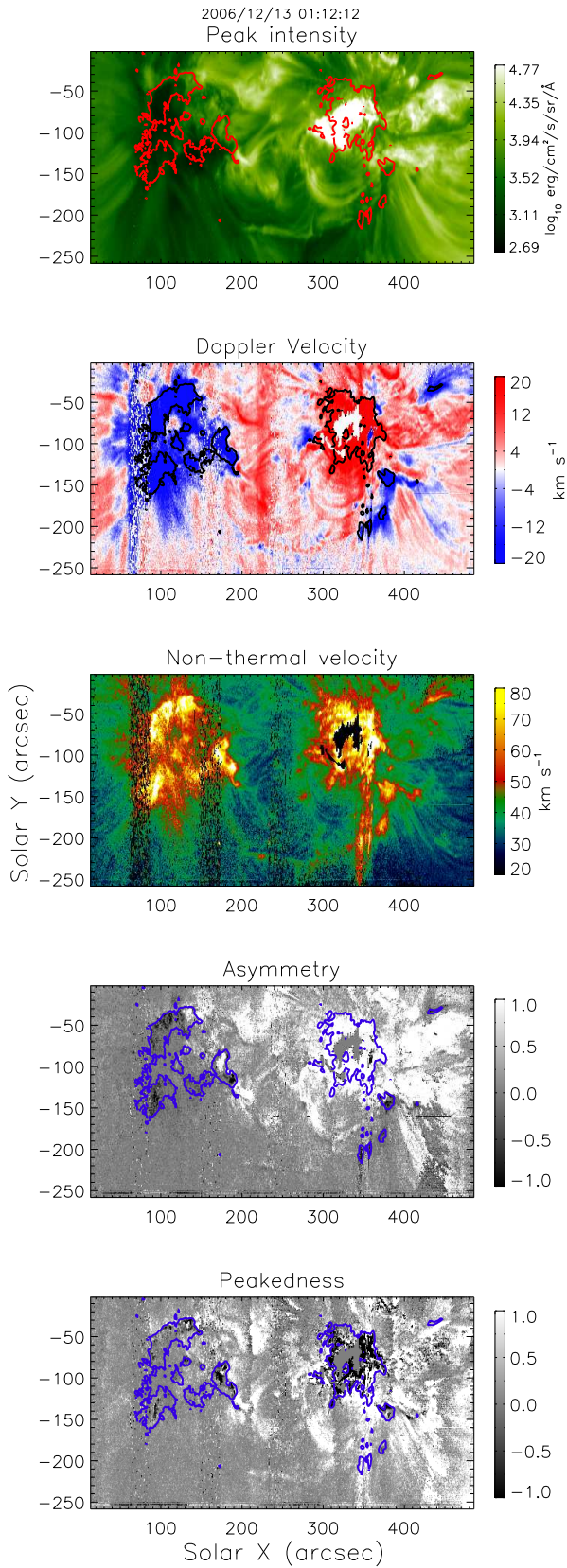}
\end{center}
\caption{Same as Fig.~\ref{fig 13 dec} but for data set~1c. 
Red, black and blue contours, in their respective panels, show the areas with a non-thermal velocity larger than $60~\textrm{km} \, \textrm{s}^{-1}$. 
\label{fig 13 dec main}}
\end{figure}
\begin{figure*}
\begin{center}
\includegraphics[height=0.95\textheight]{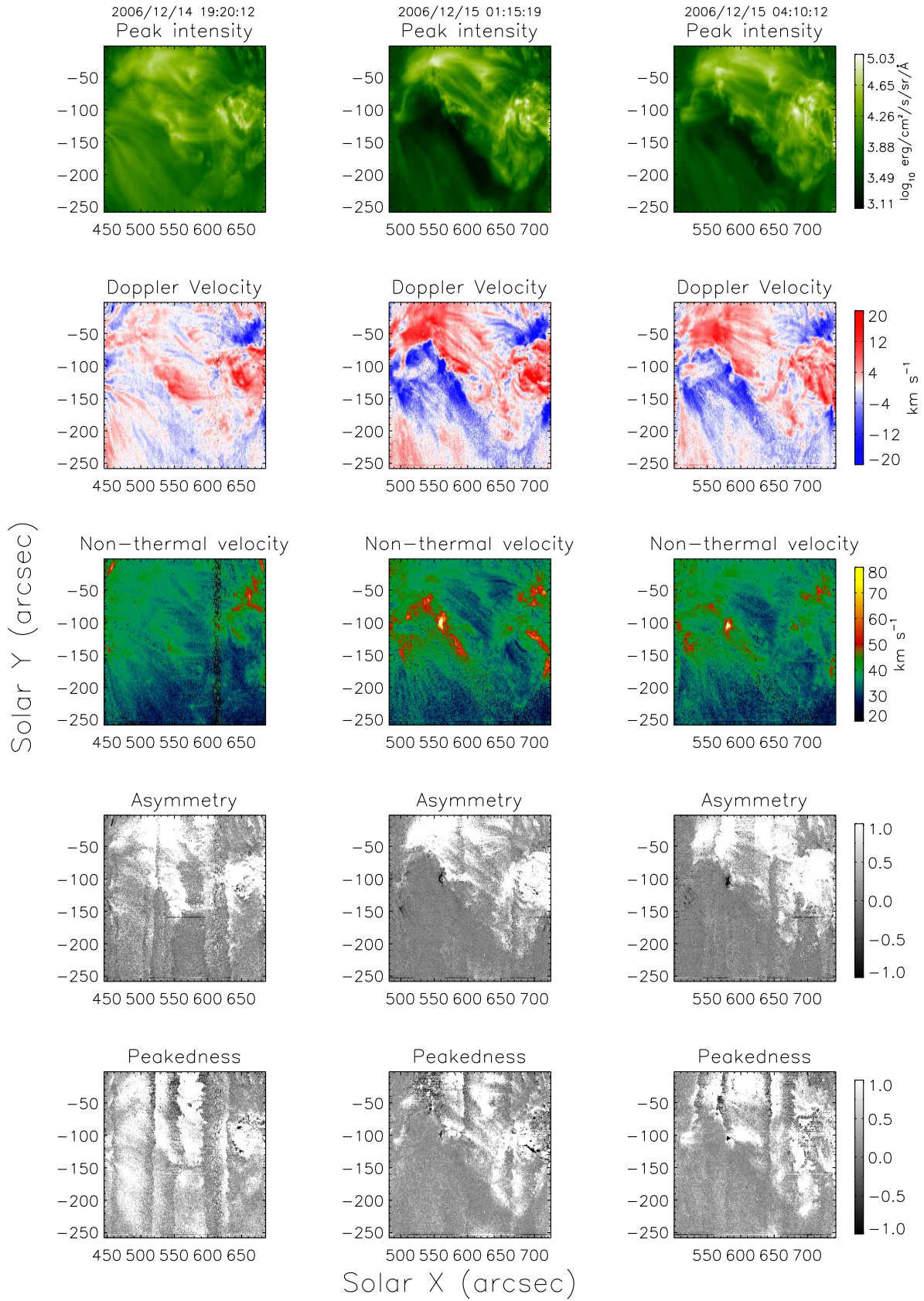}
\end{center}
\caption{Same as Fig.~\ref{fig 13 dec} but for data sets~2a, 2b, 2c, observed by \textit{Hinode}/EIS on 2006 December 14 and 15. 
\label{fig 14 15 dec}}
\end{figure*}
\begin{figure}
\begin{center}
\includegraphics[height=0.9\textheight]{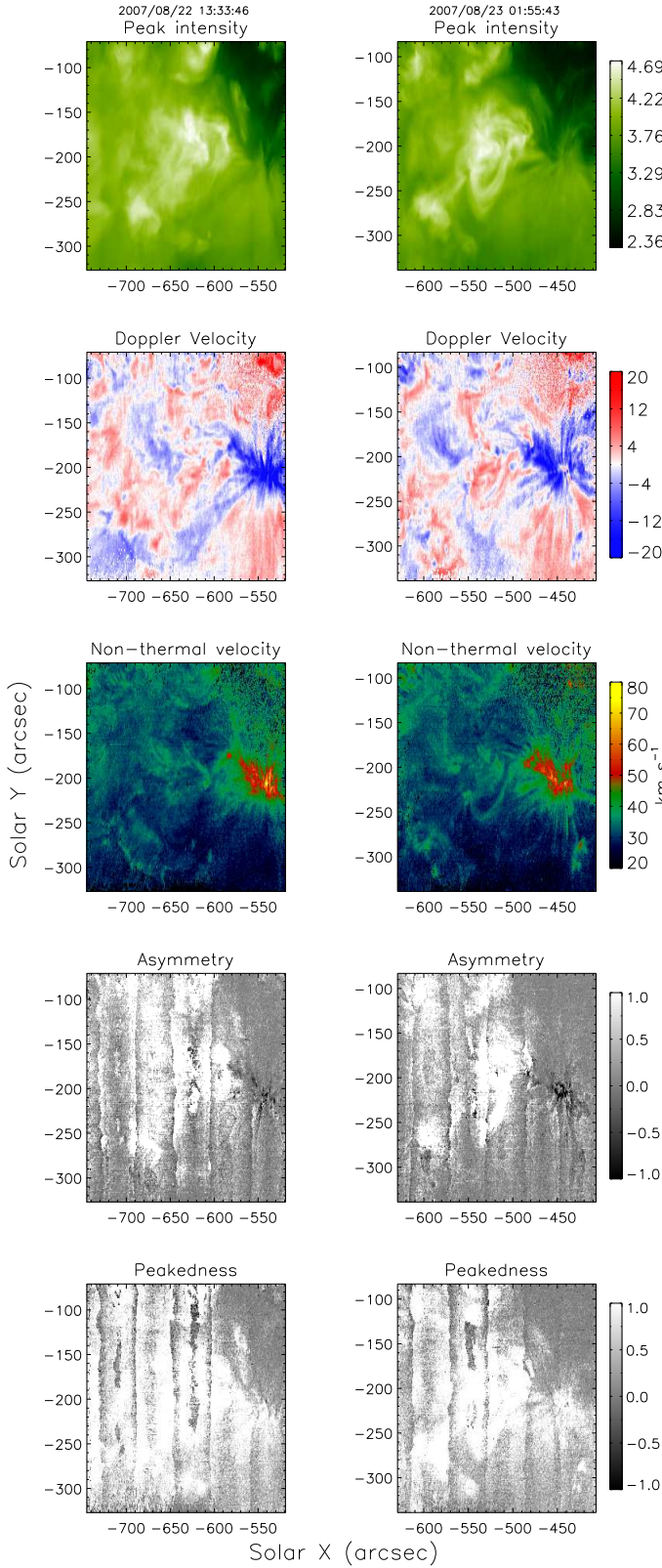}
\end{center}
\caption{Same as Fig.~\ref{fig 13 dec} but for data sets~3a and 3b, observed by \textit{Hinode}/EIS on 2007 August 22 and 23.  
\label{fig 22 23 aug}}
\end{figure}
\begin{figure*}
\begin{center}
\includegraphics[width=0.95\linewidth]{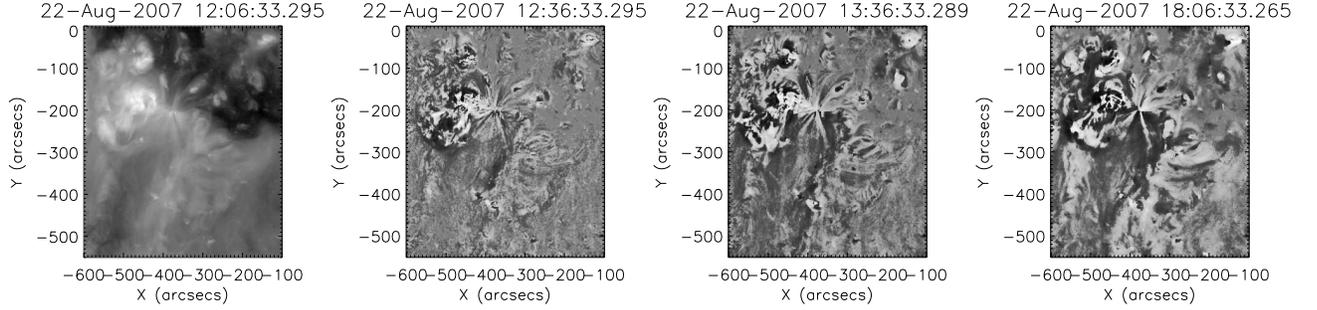}
\end{center}
\caption{Leftmost 195~\AA{} STEREO-SECCHI/EUVI panel (spacecraft B) shows the last pre-eruption image of the active region observed in data set~3. The other panels are base difference images with the image in the first panel subtracted from other images, taking into account the differential rotation. The images in the third and fourth panels were taken at the start and at the end of the raster of data set~3a. 
\label{fig diff images 22 aug}}
\end{figure*}
\begin{figure*}
\begin{center}
\includegraphics[width=0.95\linewidth]{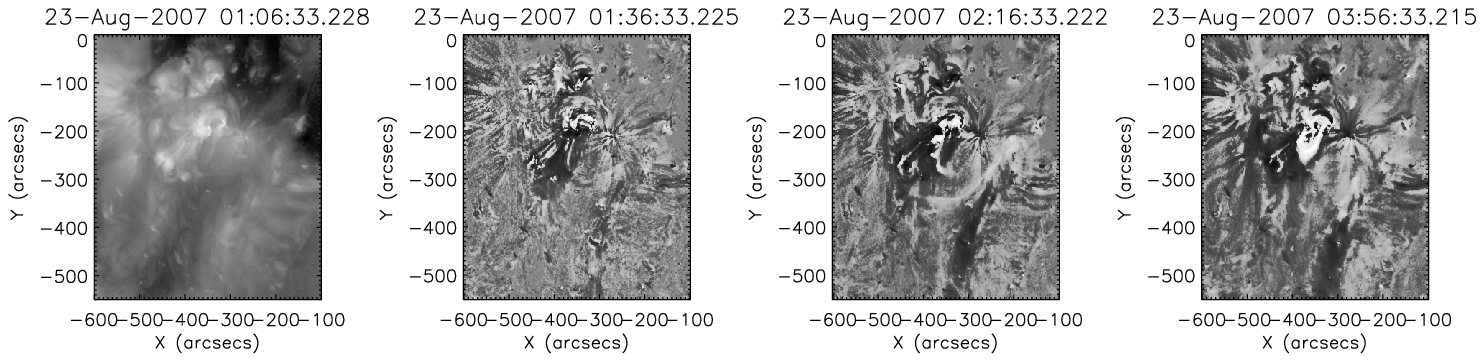}
\end{center}
\caption{The leftmost 195~\AA{} STEREO-SECCHI/EUVI panel (spacecraft B) shows the active region observed in data set~3, 50 minutes before the beginning of raster~3b. The other panels are base difference images with the image in the first panel subtracted from other images, taking into account the differential rotation. 
In the third panel, one can see dimmings and a loop system erupting. In the last panel, which corresponds to the middle of raster~3b, one can see bright loops appearing.
\label{fig diff images 23 aug}}
\end{figure*}
%
\subsection{Data processing} \label{sec data processing}
%
By fitting the line profiles with single Gaussians including a constant and a linear term, we obtain the peak intensity of the line, the LOS Doppler velocity and the non-thermal velocity derived from the line width \citep[see, e.g., ][]{Dere93}. 
For the Doppler shift, absolute wavelength calibration is difficult, so like in many other studies, we take an average wavelength position along the Y-direction as a reference, after correcting for the slit tilt and orbital variation (standard procedures included in SolarSoft). 
Although we will show that the concept of non-thermal velocity has to be taken with care, we derive it by assuming an ion temperature equal to that of maximum line formation. An instrumental width equivalent to 2.5 pixels is subtracted \citep{Doschek08}.  

When applying the \verb+eis_prep.pro+ SolarSoft procedure, we used the recommended option for interpolation of the missing data (e.g., due to the cosmic-ray removal). This procedure leaves missing data empty when interpolation has no sense (more than three neighboring  pixels in the Y-direction missing). 
If such missing data remained in the central half of the spectral window, the derived Doppler velocity, width or coefficient of distortion would be treated as missing data and set to 0 in our plots. For aesthetic reasons, the corresponding pixels in the intensity images are replaced by a median of the neighboring pixels. 

We use the pointing coordinates given in the EIS fits files: no absolute pointing precision is necessary for this study. It is well known that the pointing varies with wavelength \citep{Young07b, Young09}\footnote{see also \url{http://msslxr.mssl.ucl.ac.uk:8080/eiswiki/Wiki.jsp?page=CCDOffset} and \url{http://msslxr.mssl.ucl.ac.uk:8080/eiswiki/Wiki.jsp?page=CCDOffsetX}}. When making comparison between different lines, we realigned the images using the \ion{Fe}{12} as a reference. The misalignment in the Y-direction is corrected by using the procedure \verb+eis_ccd_offset.pro+ in SolarSoft. There is a 2-pixel difference in the X-direction between images taken with the short wavelength detector ($170-211$~\AA{}) and the long wavelength one ($246-292$~\AA{}). 
%
%
\section{Results}    \label{sec results}
As shown in Fig.~\ref{fig 13 dec} to \ref{fig 22 23 aug}, bright loops in the core of the post-flare active region present large redshifts in both legs \citep[see also ][]{Winebarger02, Delzanna08b}, while the dimming regions present large blueshifts that can exceed $50~\textrm{km} \, \textrm{s}^{-1}$ in what appears to be open magnetic field lines or very long loops. Both kinds of regions show large line widths \citep[see][]{Doschek08}. 
In set~1c (Fig.~\ref{fig 13 dec main}), the dimming area that presents large blueshifts and large widths extends out of the low-intensity part, into the quiet-Sun loops region. This area is nevertheless consistent with the area covered by the dimming in the EIT difference images \citep[see, e.g.][, their Fig. 6]{Asai08}.   
%
\subsection{Asymmetric profiles of the \ion{Fe}{12} 195~\AA{} line \label{sec asym line profiles}}
%
Regarding the asymmetry and peakedness of the \ion{Fe}{12} 195~\AA{} line, we note large areas of distorted line profiles, appearing in black or white in the last two rows in Fig.~\ref{fig 13 dec} to \ref{fig 22 23 aug}. 
The most noticeable result is that negative or positive asymmetries cluster and do not spread around randomly. The largest distortions appear in areas where the non-thermal velocity is very large. 
Asymmetries are predominantly positive in bright areas (cooling post-flare loops), while negative coefficients of asymmetry are essentially present in the low-intensity areas (dimmings). Globally, red(blue)shifted areas correspond to bright (dark) areas. 

The patterns in peakedness are less clear. Globally, bright loop areas present a positive peakedness, i.e., profiles more peaked than a pure Gaussian. There are some exceptions, like the negative areas in the core of the active region in set~1c (Fig.~\ref{fig 13 dec main}). But a closer look at the spectral profiles showed us that they correspond to large asymmetries, so that the fit is heavily shifted from an otherwise not particularly flat profile; this weights the coefficient of peakedness and shows that this coefficient can be better defined in the future. 

\begin{figure*}
\begin{center}
\includegraphics[width=0.95\linewidth]{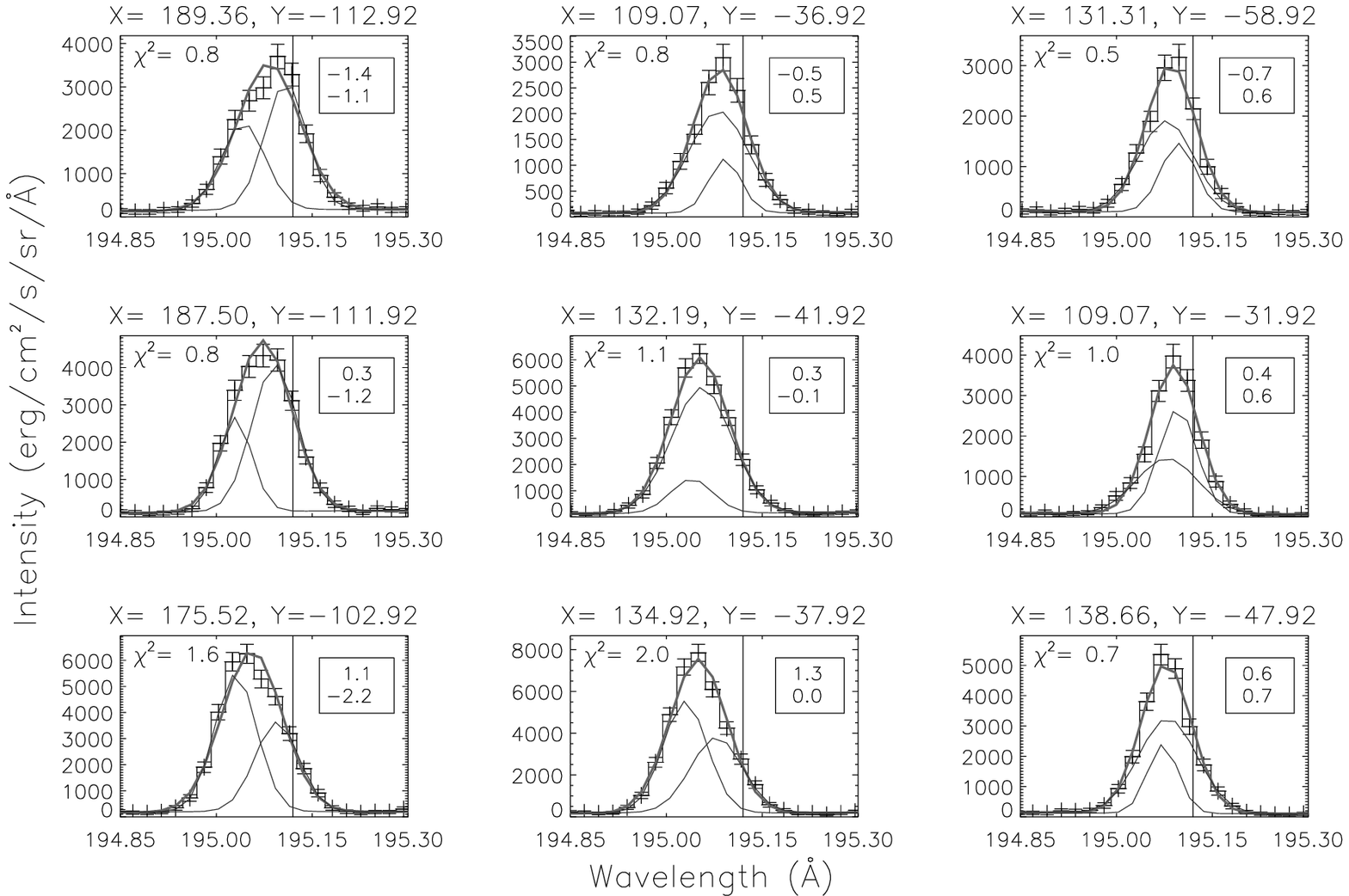}
\end{center}
\caption{Sample of \ion{Fe}{12} 195.12~\AA{} line profiles in data set~1c. We selected pixels in the dimming area (X$<250\arcsec{}$) where non-thermal velocities are larger than $60~\textrm{km} \, \textrm{s}^{-1}$ (contoured area in Fig.~\ref{fig 13 dec main}), for combinations of distortion coefficients below $-0.5$, between $-0.5$ and $0.5$, or above $0.5$. The inset in each panel contains the coefficients of asymmetry and peakedness, in this order. 
The thick gray line represents the best single-Gaussian fit to the observed profile, with the associated $\chi^2$ mentioned in each panel. Vertical lines indicate the rest wavelength. The thin gray lines represent the two components of the double-Gaussian fit. 
\label{fig sample of spectra}}
\end{figure*}

Some examples of spectra extracted from the dimming area in  data set~1c that presents large non-thermal velocity (contoured area in Fig.~\ref{fig 13 dec main}) are shown in Fig.~\ref{fig sample of spectra}, for different combinations of both coefficients. Note that in most cases, the reduced $\chi^2$ is relatively small, especially when statistics are poor, even though the distortion of the line profile can be detected by the eye. 
Our coefficients better detect the mismatch between the line profile and the fitted Gaussian.  
Following our analysis in Sect.~\ref{sec estimating line profiles}, we interpret these asymmetries as due to the presence of a minor component in the profiles: in the blue (red) wing for the negative (positive) asymmetry. This is consistent with multiple flows merging in the same LOS or pixel. 

We checked the results produced when not using the interpolation procedure for the missing data, and the asymmetry and peakedness show the same overall behavior. Therefore, the effect of the interpolation procedure for explaining the observed features can be ruled out. 

\citet{Hara08} also found, in upflows, deviations from a single Gaussian in the blue wing of the \ion{Fe}{14} 274~\AA{} and the \ion{Fe}{15} 284~\AA{} lines.  
Note that we consider here smaller outflows and much less obvious asymmetries than those presented by \citet{Imada07}, \citet{Imada08} or \citet{Asai08} for the same data set (1c), even if we also detect the prominent features that they found. In particular, the small black spot near $\textrm{X}=360\arcsec{}, \textrm{Y}=-90\arcsec{}$ in Fig.~\ref{fig 13 dec main} corresponds to the BS1 region of \citet{Asai08}. 
The method we propose enables us to emphasize more subtle distortions, and shows that they are widely present in solar active regions, including parts with downflows. 
In the contoured area of data set~1c, i.e., in pixels with non-thermal velocities larger than $60~\textrm{km} \, \textrm{s}^{-1}$  (Fig.~\ref{fig 13 dec main}), we found 21\% of pixels having an absolute coefficient of asymmetry larger than 0.5. This goes up to 28\% if we count pixels with asymmetry \emph{or peakedness} larger than 0.5. In bright loops, most of the pixels present noticeable distortion of the spectral profiles. 
%
\subsection{Double-Gaussian fit} \label{sec double Gaussian fit}
\begin{figure*}
\begin{center}
\includegraphics[width=0.95\linewidth]{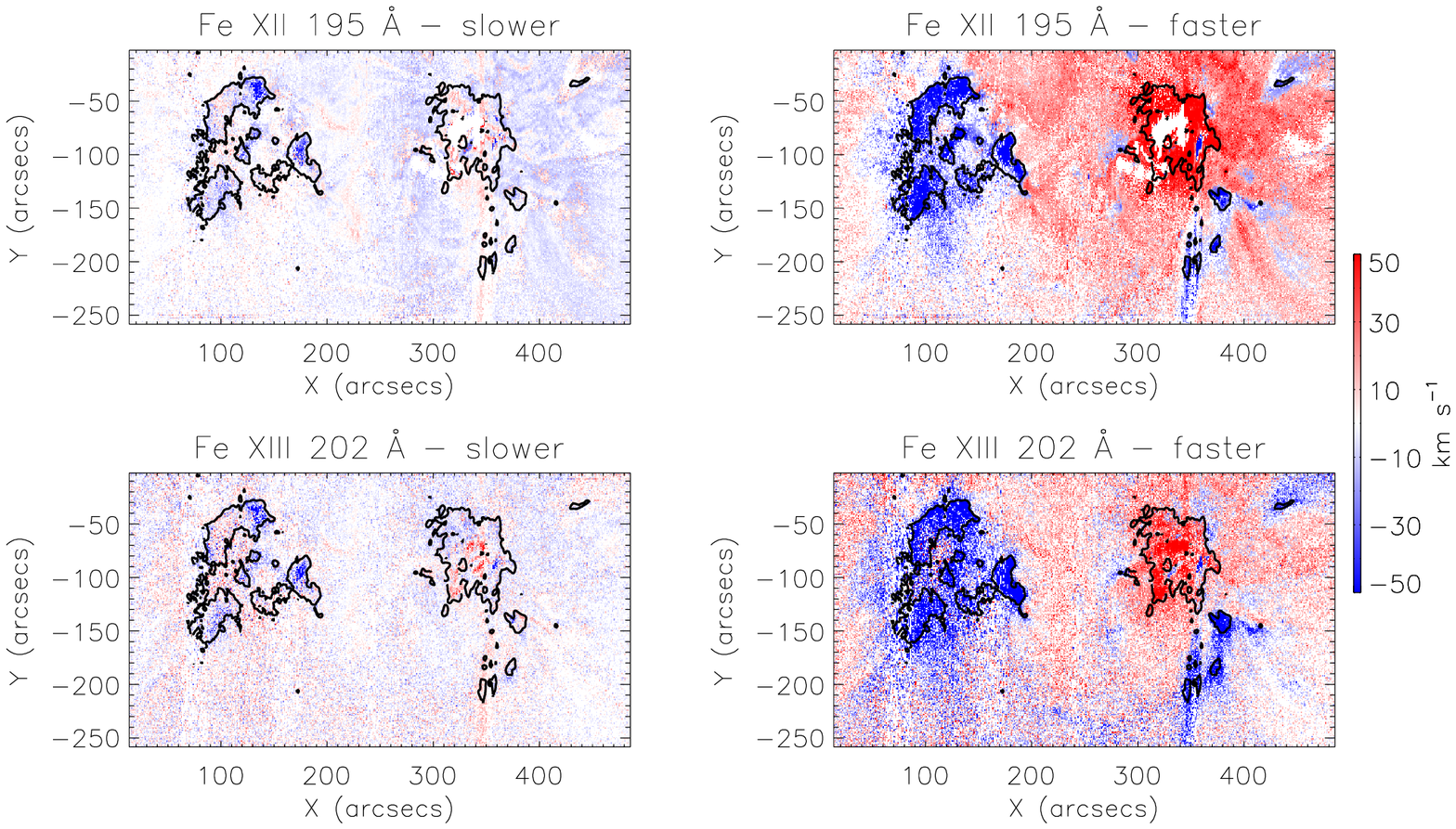}
\end{center}
\caption{Slower component (or ``nearly static'', \textit{left column}) and faster component (or ``dynamic'', \textit{right column}) of the double-Gaussian fit made for data set~1c. The \textit{top row} corresponds to the \ion{Fe}{12} 195~\AA{} line and the \textit{bottom row} to the \ion{Fe}{13} 202~\AA{} line. 
White pixels in the faster component panels mostly correspond to fits not taken into account for analysis (see the text for details).  
\label{fig Doppler double}}
\end{figure*}

%
For the multi-component analysis, we make double-Gaussian fits to the line profiles (see details in Appendix~\ref{app testing double-Gaussian fit}). 
Both components of the double-Gaussian fit for data set~1c are over-plotted in the profiles shown in Fig.~\ref{fig sample of spectra}. In Fig.~\ref{fig Doppler double} we present Doppler shifts associated with both components of the \ion{Fe}{12} 195~\AA{} line (top row) for data set~1c. 
Besides pixels with missing spectral data (see Sect.~\ref{sec data processing} and e.g., black pixels in the third panel of Fig.~\ref{fig 13 dec main}), we rejected pixels where the components are separated by more than 2 Gaussian widths of the first component. This criterion is used to prevent us from considering cases where the double fit selected neighboring lines of different rest wavelengths, that appear in the 195.12~\AA{} spectral window in some pixels. We also rejected pixels where the components are separated by less than $10~\textrm{km} \, \textrm{s}^{-1}$ (nearly one third of a spectral pixel), or where the peak intensity of one of the components is less than 20\% of the other. We applied these conservative filters to ensure that the double component that is further analyzed is significant. The white pixels in the right column of Fig.~\ref{fig Doppler double} mostly correspond to the rejected pixels. 
Judging by the retained pixels in this figure, most of the faster components are well separated from the slower ones, with a contribution of more than 20\% for the minor component, both in the dimming area and the active region core. Because of the limitations of the double-Gaussian fit explained in Appendix~\ref{app testing double-Gaussian fit}, a thorough analysis of the ratios of amplitude between the components is left for future work. 

\begin{figure*}
\begin{center}
\includegraphics[width=0.95\linewidth]{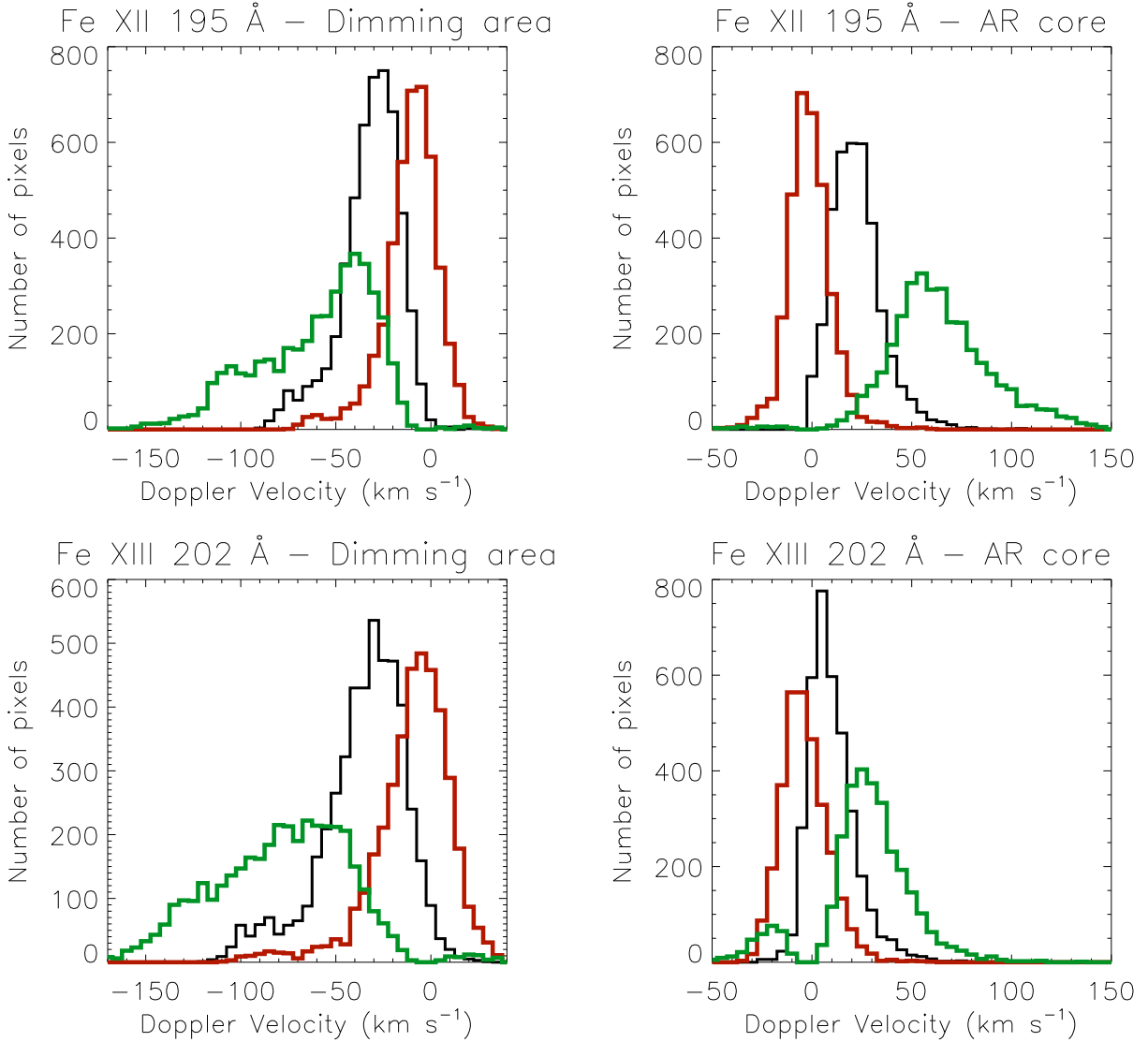}
\end{center}
\caption{Distribution of Doppler velocities measured in the pixels of data set~1c where the non-thermal velocity in the \ion{Fe}{12} 195~\AA{} line is larger than $60~\textrm{km} \, \textrm{s}^{-1}$ (calculated using a single-Gaussian fit). \textit{Left column:} for pixels in the dimming area ($\textrm{X}<250\arcsec{}$). \textit{Right column:} for the redshifted active region core (see the text). The \textit{top row} corresponds to the \ion{Fe}{12} 195~\AA{} line and the \textit{bottom row} to the \ion{Fe}{13} 202~\AA{} line. 
The black histogram corresponds to the result of the single-Gaussian fit, the red one to the static component of the double-Gaussian fit (around 0) and the green one to the dynamic component. The bin size is $5~\textrm{km} \, \textrm{s}^{-1}$. 
\label{fig distrib velocities grandes}}
\end{figure*}

The first row of Fig.~\ref{fig distrib velocities grandes} shows the distribution of Doppler velocities of both components compared to that of the single-Gaussian fit, 
for pixels where the  non-thermal velocity is larger than $60~\textrm{km} \, \textrm{s}^{-1}$ (i.e., the contoured areas in Fig.~\ref{fig 13 dec main}). 
We plot the distributions separately for pixels in the dimming area and for the active region core. In the active region core, we only retained pixels where the single-Gaussian fit provided a redshifted component. This prevents us from including the highly blueshifted event related to the flare that is studied by \citet{Asai08}, and focuses the analysis on the cooling loops. The medians of these distributions are listed in Table~\ref{tab medians velocity distributions}. 
As the error on the velocity determination 
is around a few kilometers per second, we can conclude that the slower component, in the dimming area and the active region core, corresponds to emission from a roughly static plasma. For this reason, we will hereafter call this component the \emph{(nearly) static component}. 
The second component presents Doppler velocities larger than those derived from the single-Gaussian fit. We will call it the \emph{dynamic component}. 

Regarding the non-thermal velocities retrieved from the component widths in the dimming, we find distributions with medians at 70, 42, and $76~\textrm{km} \, \textrm{s}^{-1}$ for the single-Gaussian fit, the static component, and the dynamic component, respectively. Therefore, the double component analysis shows that the static component presents a  non-thermal velocity more akin to what can be found in the quiet Sun, even though it may consist of more components, so that the individual profiles may get even narrower. 

The reliability of the double-component fit was tested on synthetic line profiles (see Appendix~\ref{app testing double-Gaussian fit}). This analysis shows that: (1) the presence of two components in the data is real and is not an artifact of simply applying the double-component model on the broad profiles; (2) the average non-zero
value of the static component is an artifact of the fitting procedure and cannot be considered as a definite result. 

Overall, the analysis we made in Sect.~\ref{sec estimating line profiles} is verified: areas with a definite positive (negative) asymmetry correspond to a redshifted (blueshifted) dynamic component in the profile. 
We also note that the redshifted dynamic component in the loop areas is larger in the core of the active region than in its surroundings. 
\begin{deluxetable*}{lcccccc}
\tablecaption{Medians of the distributions of velocities for data set~1c\tablenotemark{a}.  \label{tab medians velocity distributions}}
\tablewidth{0pt}
\tablehead{ & \multicolumn{3}{c}{Dimming area} & \multicolumn{3}{c}{Active region core\tablenotemark{b}} \\
\colhead{line ID} & \colhead{$v_\textrm{single}$\tablenotemark{c}} & \colhead{$v_\textrm{static}$\tablenotemark{d}} & \colhead{$v_\textrm{dynamic}$\tablenotemark{d}} & \colhead{$v_\textrm{single}$} & \colhead{$v_\textrm{static}$} & \colhead{$v_\textrm{dynamic}$}
}
\startdata
\ion{Fe}{12} 195.12~\AA & -27 & -6 & -50 & 24 & 1 & 64 \\
\ion{Fe}{13} 202.04~\AA & -29 & -4 & -72 & 10 & -2 & 30 \\
\enddata
\tablenotetext{a}{corresponding to the contoured areas in Fig.~\ref{fig Doppler double}, i.e., non-thermal velocities larger than $60~\textrm{km} \, \textrm{s}^{-1}$ with the single-Gaussian fit. All velocities are in $\textrm{km} \, \textrm{s}^{-1}$. }
\tablenotetext{b}{For the active region core, only pixels with redshifted single-Gaussian component are retained (see the text). }
\tablenotetext{c}{Derived from the single-Gaussian fit. }
\tablenotetext{d}{Derived from the double-Gaussian fit. }
\end{deluxetable*}
%
%
\subsection{Results for other spectral lines} \label{sec different lines}
%
Several other lines were recorded in our data sets. We concentrate again on data set~1c that shows the asymmetric profiles more clearly. As the signal in other lines is lower, we spatially rebinned the data by 3~pixels in the X-direction and 3~pixels in the Y-direction. We verified that this rebinning does not produce additional distortions (Appendix~\ref{app effect rebinning distortion}). 
This also demonstrates that the source of the distortion has nothing to do with the resolution scale: it has to be found in the LOS integration, in scales much smaller than the resolution scale or in temporal variation. 

\begin{figure*}
\begin{center}
\includegraphics[width=0.9\linewidth]{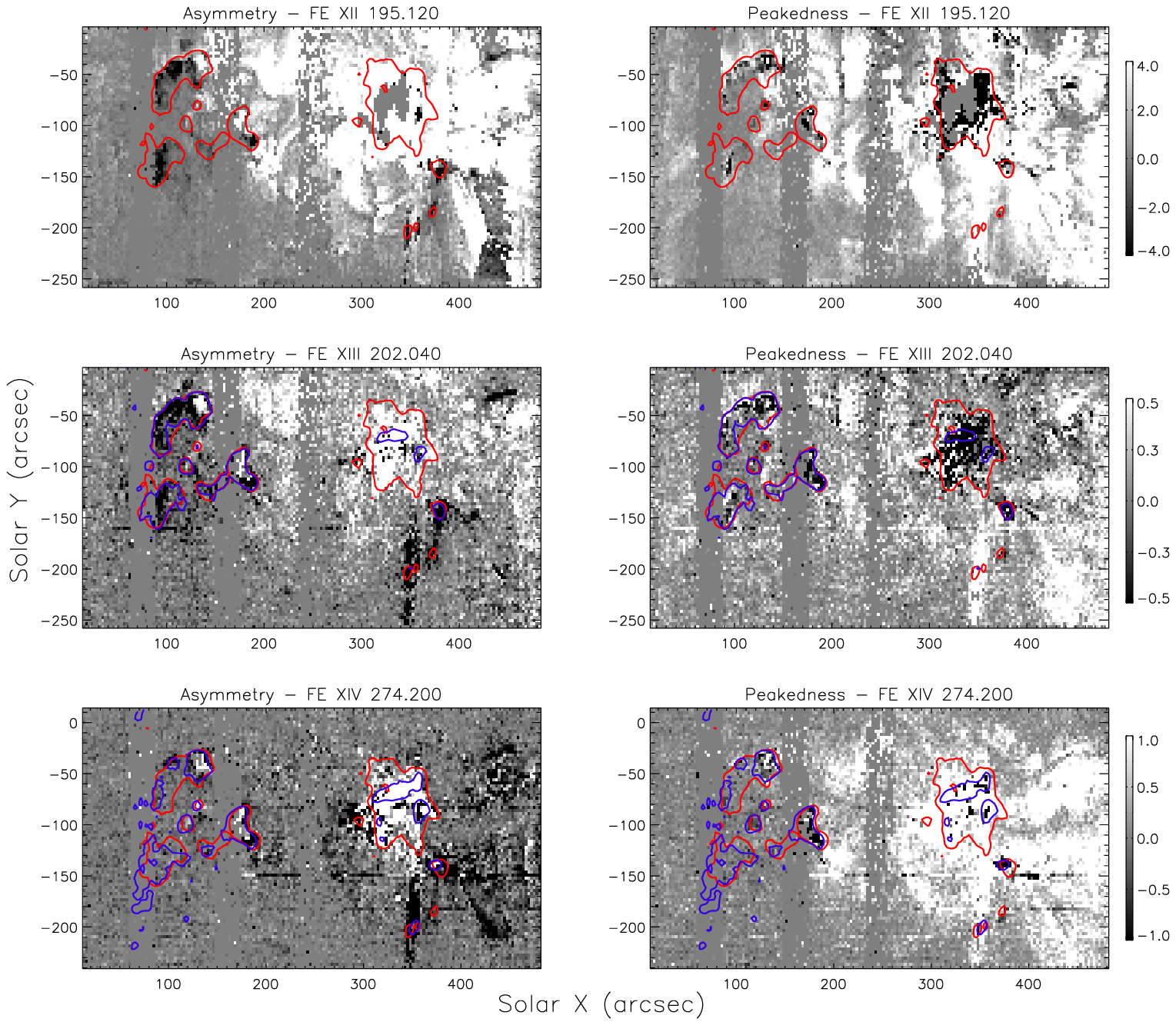}
\end{center}
\caption{Coefficients of asymmetry (left column) and peakedness (right column) for the \ion{Fe}{12} 195.12~\AA{}, \ion{Fe}{13} 202.04~\AA{} and \ion{Fe}{14} 274.20~\AA{} lines. EIS pixels are binned by 3 in each direction. Red contours correspond to $60~\textrm{km} \, \textrm{s}^{-1}$ for the non-thermal velocity of \ion{Fe}{12}, while the blue contours correspond to the same limit for each of the two other lines. 
\label{fig asym peak other lines}}
\end{figure*}

After the rebinning, we derived the coefficient of asymmetry and peakedness for several lines: \ion{Fe}{12} 195.12~\AA{}, \ion{Fe}{13} 202.04~\AA{} and \ion{Fe}{14} 274.20~\AA{} (see Fig.~\ref{fig asym peak other lines}). These ions have peak formation temperature increasing from 1.6 to 2~MK. 
As there is a wavelength-dependent shift of the spatial position on the detector, we co-aligned all maps with that of the \ion{Fe}{12} line (see Sect.~\ref{sec data processing}). 
To ease the comparison, we overplot the $60~\textrm{km} \, \textrm{s}^{-1}$ contours for the non-thermal velocity of \ion{Fe}{12} ions on all the panels. Different thresholds for the color scale are used to ease the visual comparison. It must be emphasized that the differences in the signal statistics greatly influence the value of the coefficients due to the inclusion of the error bars in Eq.~\ref{eq coeff}. 

Areas of large line widths (blue contours) appear more fragmented when the formation temperature increases, but this may be a selection effect due to the definition of the non-thermal velocity as the excess width as compared to the formation temperature of the line. 
The contours are mainly co-spatial for different lines in the dimming area. We also note that the same areas of negative or positive asymmetry or peakedness appear at the same locations for every wavelength, especially in the dimming area for \ion{Fe}{12} and \ion{Fe}{13}. We do not show the Doppler and non-thermal velocity maps, but they are similar to those taken in the \ion{Fe}{12} line, as has already been shown in several studies \citep[e.g.][]{Imada07, Jin09}. No large distortions are found in the dark pixels of the central dimming area for \ion{Fe}{14}, which is essentially due to the lack of photon statistics there. 

Among the other available lines, the \ion{Fe}{8} 185.21~\AA{} line is too faint and leads to very noisy maps of asymmetry, peakedness and even line widths. For the \ion{Fe}{10} 184.54~\AA{} line, we find distorted profiles in the active region, but the coefficients of distortion are close to 0 in the dimming area, even though we could discern some black patches reminiscent of what we find in other wavelengths. Again, this may be a question of lack of photon statistics. The \ion{Fe}{11} 188.23~\AA{} line is composed of two close lines of the same ion. This would make the distortion analysis difficult. The analysis of the \ion{Fe}{15} 284.16~\AA{} line is complicated by blending issues. 

We also find a similar behavior for data set~3. Positive (negative) asymmetry is seen in the loop (fanning-out structure) area, for the same lines as shown in Fig.~\ref{fig asym peak other lines} and for additional lines, including another \ion{Fe}{12} line at 193.51~\AA{} and another \ion{Fe}{14} line at 264.78~\AA{} (not shown). 

As demonstrated in Appendix~\ref{app testing double-Gaussian fit}, the double-Gaussian fit of closely situated lines is not precise enough to allow a reliable comparison pixel by pixel. This is especially true for different wavelengths, for which one has also to correct accurately pixel shifts on the detector due to the wavelength dependence. This effect is even more important when the lines are recorded on the different EIS detectors. We show the Doppler maps and velocity distributions for the \ion{Fe}{13} 202~\AA{} line in Fig.~\ref{fig Doppler double} and Fig.~\ref{fig distrib velocities grandes} respectively. 
%
%
\section{Discussion}
%
We found a clear pattern of asymmetry of coronal line profiles. In dimmings, a substantial part of them is left-skewed (negative asymmetry), while in bright loops most of them are right-skewed (positive asymmetry).  The largest distortions appear in areas where the non-thermal velocity or, more precisely, the width of the profile, is very large. This is especially true in the left-skewed profiles. A double-Gaussian fit reveals that these asymmetric profiles are composed of one nearly static component and (at least) one dynamic component. Most of the time the dynamic component is systematically blueshifted (redshifted) in the case of a left-skewed (right-skewed) profile.
%
\subsection{Alfv\'en waves or flow inhomogeneities ?}
It appears that transverse velocity components of MHD waves and, in particular, Alfv\'en waves as suggested by \citet{McIntosh09} cannot be responsible for the line broadening observed in dimmings. If this were true, the broadened profiles should be symmetric, because velocity perturbation in a wave oscillates around zero. If one assumes that the spectrometer takes a snapshot of MHD waves having periods longer than the exposure time (30~s per X-position), one should find an approximately equal number of points with negative or positive asymmetry, randomly positioned. One could also imagine that all the waves in adjacent pixels are in phase so that their velocity perturbations all have the same sign. But this would have to be the case not only in the Y-direction (simultaneous exposure) but also in the X-direction, which implies the coincidence of the rastering speed with the wave propagation speed. This appears very unlikely, especially in so many different data sets. Above all, waves cannot explain the fact that the left-skewed profiles are found in dimmings and the right-skewed profiles in loops. 

Our interpretation appears more plausible: the broadened profiles are due to the additional component that has the same velocity direction as that found with the single-Gaussian fit. 
Downflows in loops may correspond to draining of cooling material \citep[e.g.][]{Bradshaw08}, while upflows correspond to  material escaping the corona or refilling it after the eruption. 
The multi-component approach offers a new point of view on the ``coronal circulation'' concept suggested by \citet{Marsch08}: the dynamic component is most of the time observed together with a nearly static one in the same LOS.  
Plasma velocity varying quickly on temporal scales smaller than the exposure time (associated with jets or any other short duration events), may be another explanation. 
It is noteworthy that these interpretations are compatible with the fact that a lot of line profiles do not show any substantial distortion in the dimming areas that nevertheless present a broad profile (e.g., the central panel of Fig.~\ref{fig sample of spectra}). 
Indeed, for small differences in the centroid positions of the components or for small relative intensity of the additional component, the resulting profile is broadened but does not present a noticeable distortion, like in Fig.~\ref{fig simulated profiles}~\textit{(a)}. In this simple two-component case, the width of the resulting profile is increased by about $10\%$ as compared to that of the single components. This turns into an increase of 30\% in non-thermal velocity in the case of an \ion{Fe}{12} line. This value is larger than the average increase observed in the dimming area of data set~2 as shown in Fig.~\ref{fig 14 15 dec} \citep[see also panel (D) in the Figure 3 of the work by][]{McIntosh09}. 
The lack of photon statistics in dimmings may be another reason for non-detected distortions. But the additional component is probably present in all pixels with large line profiles, and thus can explain the observed increase of the coronal line widths. 
%
\subsection{Effect of line blending} \label{sec line blending}
We now assess the effect of line blending on the observed distortions. 
First of all, let us remind that to calculate the coefficients of asymmetry we select only a window around $2 \sigma$ from the line center (see Eq.~\ref{Eq intervals asym}). This means that one can consider any line lying beyond that interval as not contributing to this coefficient. The interval is even smaller for the peakedness (see Eq.~\ref{Eq intervals peak}). Note also that when considering the double-Gaussian analysis, we discard any fits where the center of the second component is beyond $2 \sigma$ from the main component. This filter is applied after all the iterations of the fitting procedure (not during the iterations), which means that we have no selection effect. Note also that a very few of such fits were discarded. They are located mainly in the active region near X=300\arcsec{}, Y=-120\arcsec{} in the case of \ion{Fe}{12} (see Fig.~\ref{fig Doppler double}, where they appear as the white pixels that do not correspond to the black pixels of the line-width panel in Fig.~\ref{fig 13 dec main}). 

A $2 \sigma$ interval for the \ion{Fe}{12} 195.12~\AA{} line makes on average 0.06~\AA{} (or about $92 ~\textrm{km} \, \textrm{s}^{-1}$). It can go up to 0.1~\AA{} in the large widths areas (dimming and active region core). \citet{Brown08} report on the following neighboring lines for \ion{Fe}{12} 195.12~\AA{}. There is an \ion{Fe}{12} 194.92~\AA{} line in the red wing. This is more than $4 \sigma$ apart (and it is only less than 2\% of the 195.12~\AA{} line, according to these authors). On the blue wing, the closer one is \ion{Fe}{14} 195.25~\AA{}, which is more than $4 \sigma$ apart from the 195.12~\AA{} line, or $2.5 \sigma$ if we consider the large width case (and again, it is less than 2\% of the intensity of the \ion{Fe}{12} line). Thus, this line is not taken into account in our computations and analysis. In the CHIANTI database \citep{Dere97, Dere09}, we also find a \ion{Ni}{16} line at 195.27~\AA{} which is again too far. The only critical blending issue is another \ion{Fe}{12} line at 195.18~\AA{}. 
For a density of $10^{10}~\textrm{cm}^{-3}$, i.e in active regions, it is predicted to contribute around 10\% in intensity \citep{DelZanna05}. When we simulated such kind of line profiles with the high photon statistics characteristic of active regions, we found coefficients of asymmetry larger than 1. But the coefficients really measured in data set~1c are much larger. Moreover, the distortion is hardly distinguishable by the eye in the simulated profiles, contrary to that observed in the real active region profiles. This suggests that the main contribution for the positive asymmetry of the \ion{Fe}{12} 195.12~\AA{} line profile in loop areas is not the self-blend with \ion{Fe}{12} 195.18~\AA{}. The self-blend cannot account anyway for the negative asymmetry in the dimming.
 
According to \citet{Young07}, the \ion{Fe}{13} 202.04~\AA{} line is unblended, the \ion{Fe}{14} 274.20~\AA{} line blend with a \ion{Si}{7} line (in the blue wing) can be neglected in active regions \citep{Liang10}. 

Therefore, the negative asymmetry in dimmings, free of any blending, can be ascertained from the \ion{Fe}{12} 195.12~\AA{} 
and \ion{Fe}{13} 202.04~\AA{} lines. The positive asymmetry in loops, at least in the active region core, can be ascertained from the \ion{Fe}{13} 202.04~\AA{}. This positive asymmetry can also be ascertained from the \ion{Fe}{14} 274.20~\AA{} line because the weak blend it is subject to is present in the blue wing, which means it should produce a negative, and not positive asymmetry. 
Overall, similar patterns of asymmetry and/or peakedness observed in different lines (see Fig.~\ref{fig asym peak other lines}) suggest that the line profile distortions are not essentially produced by blending. 
Inhomogeneities of flow then appear as the main cause of these asymmetries, even though line blending complicates the analysis of the components. 
%
\subsection{Flow geometry and link with eruptions}
%
We showed that rebinning the data does not produce additional distortion of line profiles. On the contrary, it helps to emphasize the pre-existing distortions. Consequently, the inhomogeneities in velocities on scales larger than the resolution scale are not sufficient to produce the observed distortions: such inhomogeneities must be already present before the binning. They should exist on scales smaller than the initial resolution scale ($\approx 1\arcsec{}$), or must be due to LOS or temporal integration. 

An interesting question is the nature of the static component, most of the time the more intense one. It could correspond to a zero sum between blue and red shifts that will symmetrically broaden the static component in excess of the thermal width (the usual interpretation of the non-thermal velocity). This can be caused by waves or turbulence, for example Alfv\'en waves. 
Another possible interpretation is the presence of flux tubes where the plasma has no or small overall projected velocity on the LOS. In addition to a truly static component, this may also be the case when the LOS (or pixel) superposition of many flux tubes with different (counter-streaming) flow velocities is present. Additionally, the cross section of some of the flux tubes may be perpendicular to the LOS (e.g., near the loop apexes).
These interpretations can hold both for loops (multiple parallel strands in one loop or overlying different loops) or the dimming areas. In the latter case, the areas of negative asymmetry seem to present a radial pattern (especially for data sets 1b, 1c, 3a and 3b, see the black areas in the corresponding panels in Fig.~\ref{fig 13 dec}, \ref{fig 13 dec main} and \ref{fig 22 23 aug}). It is tempting to link them to flux tubes of the fanning-out structure. The static component could then correspond, here again, to neighboring radial strands with plasma at rest or to apexes of very long loops overlying the dimming area. 
In three dimensions, one can imagine a mix of field lines, some of them nearly vertical and maybe open, some other ones more inclined or even horizontal, all more or less entangled, with proportions depending on the local topology of the magnetic field. 

Note also that more components may be present in the line profiles. Due to their closeness, it is very difficult to distinguish them. The profiles do not present enough ``bumps'' to  unambiguously retrieve every component by a multi-component fitting procedure. With the present analysis, the only thing we can be certain of is that these profiles are not symmetric enough to be explained by a Gaussian or any other symmetric distribution of velocities for the components. There are at least two components. 

It is tempting to relate the systematic blueshift of the nearly static component to a real flow corresponding, e.g., to the nascent solar wind. But first, one has to remember that this Doppler velocity has been derived by taking a reference wavelength computed by using the centers of the single-Gaussian fits, i.e., not taking into account the fact that some of these profiles have multiple components. Second, this systematic shift may be a bias of the two-Gaussian fit (see Appendix~\ref{app testing double-Gaussian fit}). Third, these Doppler shifts are within the error bars. 
This prevents us from firmly concluding on the reality of this upflow. 

Another interesting issue is the relation between negative asymmetries and eruptions. In our data sets, it is obvious that the negative asymmetries get larger and more extended after every eruption (cf Sect.~\ref{sec obs and data processing} and Fig.~\ref{fig 13 dec} to \ref{fig 22 23 aug}). 
We note that data sets 1 and 2 are taken in the same active region that underwent a significant rotation from near the disk center toward the west limb. Nevertheless, the Doppler shift patterns as shown in Fig.~\ref{fig 13 dec}, \ref{fig 13 dec main} and \ref{fig 14 15 dec} look very similar. We could not find any marked difference between the flows due to the different positions on the disk. 

Three effects can explain the increase of asymmetries after eruptions. First, we can suppose that the high-speed dynamic component is already present before the eruption. 
The disappearance of a large part of the coronal material, especially in the apexes of closed loops discussed above, may lower the intensity of the static component and thus make the dynamic one more prominent in the profile. This can be the case for data sets~1 and 2. This interpretation is reinforced by the fact that between data sets~1c and 2a, the transient coronal hole has been refilled, while the negative asymmetries have disappeared. 
The case of data set~3 is more complex: the observed dimmings here are very faint and do not really correspond to the most blueshifted structure. 

Still with a pre-existing dynamic component, a second explanation is that the reorganization of the magnetic topology during the eruption leads to a better alignment of the flows with the LOS (e.g., more vertical flux tubes, as expected during a dimming event). 
This increases the LOS component of the flow and the asymmetry of the resulting profiles. This explanation can account for the transition between data sets~3a and 3b, or maybe for the transition from data set~2a to 2b. 
But there is a large area of redshifts in data set~1b around  $\textrm{X}=50\arcsec{}, \textrm{Y}=-80\arcsec{}$. 
After the eruption, these areas show only blue shifts. This implies that the flows reverted from downward to upward. It is then difficult to envisage that all the outflows were present before. 

The third possibility is that the dynamic component may be a feature appearing in dimming only after eruptions. 
It is possible that dimmings are magnetically open structures \citep[see e.g.][]{Kahler01, Attrill08, McIntosh09}. The plasma then can easily escape from the Sun, producing the observed outflows \citep[e.g.][]{McIntosh10}. 
From this point of view, it is possible that apparently broader profiles (cf Sect.~\ref{sec double Gaussian fit}) of the dynamic component correspond to the acceleration that produces sub-components of increasing blue shift during the exposure time. They would then be considered as a single dynamic component by the double-Gaussian fit. 
Therefore, the increase of asymmetry of the coronal line profiles may be due to an increase of the transient activity after the eruption. Whatever the case may be, the outflow corresponding to the dynamic component may take part in the refilling of the corona during the dimming recovery. 

It is worthwhile to emphasize that most of the LOS velocities that we find for the dynamic component are subsonic (the sound speed $c_\textrm{s} \approx 200~\textrm{km} \, \textrm{s}^{-1}$ at $1.6 \times 10^6$~K; cf Fig.~\ref{fig distrib velocities grandes}). 
%
%
\section{Conclusions}   
We analyzed the profiles of coronal iron lines observed with \textit{Hinode}/EIS by using empirical coefficients (asymmetry and peakedness) efficient at both detecting and discriminating different distortions of the line profiles. These coefficients are sensitive to the photon statistics in the profiles, but this is what makes them reliable. They are a good complement to multi-component analysis, particularly for a preliminary check of interesting data sets, as they are fast to compute. They are especially suitable  to reveal the presence of very close components, which are difficult to resolve with a double-Gaussian fit. 

We showed that the line broadenings correlated with large Doppler shifts observed in several active regions, 
are associated with the presence of more than one Gaussian component in the line profile. Upflows are found in the coronal dimming area and downflows in the loops, confirming previous studies \citep[e.g.][]{Harra01, Winebarger02}.  However, their velocities are underestimated if a single-Gaussian fit for the line profile is used. Both static plasma and large flows (several tens of kilometers per second) are present in the same LOS or pixel. 

We do not rule out the interpretation that the inherent density decrease in coronal dimmings can broaden the line profiles due to the induced increase of the Alfv\'en wave amplitude, as suggested by \citet{McIntosh09}, 
but this cannot be the sole cause. It is difficult to explain the asymmetries observed on large scales only by Alfv\'en waves. One has to take into account the effect of inhomogeneities of velocities in the LOS (overlapping strands with different inclination or intrinsic velocity) or in the spatial resolution pixel (flows with sub-resolution structure), or transient variations of the velocity on timescales smaller than the exposure time (30~s). 
The concept of non-thermal velocity derived from a single-Gaussian fit must then be handled with care. In particular for on-disk observations, one cannot directly interpret this quantity as the measure of average fluctuations of velocities in the observed pixel, especially in terms of Alfv\'en wave amplitude. 
We demonstrated that this pattern is visible in three types of coronal structures which exhibit large-scale flows: CME-associated dimmings, loops, and fanning-out structures. 
A new field of investigation is now opened to analyze the structure of inhomogeneities in an optically thin coronal medium. 

Finally, we note that since the submission of this paper, three more works emphasized the importance of asymmetric coronal line profiles. \citet{McIntosh10} use a different approach based on the method described by \citet{McIntosh09c} to analyze the \ion{Fe}{13} 202~\AA{} and \ion{Fe}{14} 274~\AA{} lines. 
They redress the Alfv\'en wave interpretation suggested by \citet{McIntosh09} and independently arrive at the same conclusion as reported in our work: the analysis of line profile asymmetries shows that multiple component flows (and not only Alfv\'en waves) play an important role in the plasma dynamics in coronal dimmings. 
\citet{Peter10} finds in an active region a narrow line core and a broader minor component blueshifted by up to $50 ~\textrm{km} \, \textrm{s}^{-1}$ for the \ion{Fe}{15} 284~\AA{} line.  
\citet{Bryans10} showed that, in outflows observed near an active region, the \ion{Fe}{12} and \ion{Fe}{13} line profiles are better represented by double-Gaussian fits. We find this behavior not only in upflows associated with dimmings events, but also in downflows associated with loop structures. 
\acknowledgments
We thank the referee, S. McIntosh, for his useful suggestions, in particular to present more data sets and wavelengths to ascertain the observed asymmetries. \textit{Hinode} is a Japanese mission developed and launched by ISAS/JAXA, with NAOJ as a domestic partner and NASA and STFC (UK) as international partners. It is operated by these agencies in co-operation with ESA and NSC (Norway). CHIANTI is a collaborative project involving researchers at NRL (USA) RAL (UK), and the Universities of: Cambridge (UK), George Mason (USA), and Florence (Italy). 

\emph{Facilities:} \facility{Hinode (EIS)}, \facility{STEREO (SECCHI/EUVI)}, \facility{GOES}
\appendix
\section{Double-Gaussian fit and tests on synthetic line profiles} \label{app testing double-Gaussian fit}
We used a double-Gaussian fit procedure including a constant and a linear term, based on a random-restart hill-climbing method. Basically, a classical gradient-expansion algorithm (\verb+CURVEFIT.pro+ in IDL) computes a non linear least-squares fit. It is restarted more than $10^4$~times, finally selecting the solution presenting the best $\chi^2$. We used random starting estimates that could vary relatively widely: peak intensity between 0 and the maximum of the spectrum in the recorded window, line center within the wavelength interval of this window and line width between the instrumental width and $1/4$ of the spectral window width. To be selected, the solution of the gradient expansion had to keep into those limits. Hereafter, one has to understand the solution of a ``double-Gaussian fit'' as the best solution of multiple iterations. 

We tested the reliability of this procedure on synthetic line profiles. First we simulated a double-Gaussian profile with components similar to what we find in the contoured area of the dimming in data set~1c: a component at rest and an additional one with 40\% amplitude, offset by a few tens of kilometers per second, but having the same width as the static component. We applied the double-Gaussian fit thousands of times on the same synthetic profile (i.e., each solution of a fit being itself the results of thousands of iterations). 
We then analyzed the resulting statistics. 
Even without added statistical noise on the synthetic profiles, the amplitude, velocity and line width of the individual components cannot be retrieved with high precision.  
Not surprisingly, the parameters of the component having the lowest amplitude present the largest dispersion. Its amplitude is on average overestimated by 25\%. The distribution of retrieved component centers is rather flat over an interval more than $60 ~\textrm{km} \, \textrm{s}^{-1}$ wide, with a median displaced by about $25~\textrm{km} \, \textrm{s}^{-1}$ as compared to the initial value. This means that the center of the small (and dynamic, in our case) component is on average artificially found redshifted as compared to the original component. We remind that an EIS spectral pixel is equivalent to $\approx 35~\textrm{km} \, \textrm{s}^{-1}$. Likewise, the line width of the small component is on average overestimated by 15\%. In compensation, the center of the static (and more intense) component is artificially blueshifted by about $10~\textrm{km} \, \textrm{s}^{-1}$, which is similar to what we find in real data. When we add statistical noise, the results are not modified very much. This analysis holds for profiles where no clear double-peak appears in the resulting profile; it is of course easier to retrieve the correct parameters when the two components are well separated. 

In a second kind of verification, we simulated a single, large Gaussian and try to fit it with two Gaussians. Not surprisingly, when no noise is added, the fit provided two nearly identical Gaussians, with an amplitude half of the original one. When we add some statistical noise, the fitting procedure starts to ``lock'' on some local bumps on the profile and finds two components that can be separated by up to several tens of kilometers per second. On average, they nevertheless remain separated only by about $\pm 10~\textrm{km} \, \textrm{s}^{-1}$. 

All these results correspond to a few particular sets of parameters and cannot be compared directly to our data set that more likely corresponds to a wider range of parameters. But it gives us important information about the reliability of results of our double-Gaussian fit: the presence of two components (at least) is real in the pixels with large non-thermal velocity. The fact that one of them is at rest and the other one is highly Doppler-shifted is statistically relevant, even though it is impossible to have confidence in the fit parameters for a given pixel. Our analysis also shows that one should be careful in interpreting the results of our double-Gaussian fit for pixels with small non-thermal velocity.  
Only when the coefficient of asymmetry and/or peakedness are non-negligible, one can trust the double-Gaussian fit. 
%
\section{Effect of rebinning on measured distortions} \label{app effect rebinning distortion}
Due to the orbital variation of the line center and to the spatial inhomogeneity of the Doppler shift, the rebinning (cf Sect.~\ref{sec different lines}) can be an additional source of distortion for the line profiles. We have to verify, then, that the rebinning does not contribute much to the distortion we want to emphasize. For that, we simulated a $3 \times 3$ rebinning but with the single-Gaussian fitted profiles of \ion{Fe}{12} instead of the real data. The derived maps of asymmetry and peakedness (not shown) present no particular pattern, and are absolutely not comparable to what we obtained in Fig.~\ref{fig 13 dec main} without rebinning. In fact, the coefficients are nearly equal to 0 for almost all pixels. 
We can then conclude that the rebinning does not have any effect on revealing the distortion of the line profiles qualitatively. This also demonstrates that the source of the distortion has nothing to do with the resolution scales: it has to be found in the LOS integration, in scales much smaller than the resolution scale or in temporal variation. 
\end{document}